# Simulation of White Light Generation and Near Light Bullets Using a Novel Numerical Technique

Haider Zia[1]*

[1]*Max-Planck Institute for the Structure and Dynamics of Matter, Luruper Chausee 149, Hamburg, DE-22607*
*\*haider.zia@mpsd.mpg.de*

**Abstract:** An accurate and efficient simulation has been devised, employing a new numerical technique to simulate the derivative generalised non-linear Schrödinger equation in all three spatial dimensions and time. The simulation models all pertinent effects such as self-steepening and plasma for the non-linear propagation of ultrafast optical radiation in bulk material. Simulation results are compared to published experimental spectral data of an example ytterbium aluminum garnet system at 3.1µm radiation and fits to within a factor of 5. The simulation shows that there is a stability point near the end of the 2 mm crystal where a quasi-light bullet (spatial temporal soliton) is present. Within this region, the pulse is collimated at a reduced diameter (factor of ~2) and there exists a near temporal soliton at the spatial center. The temporal intensity within this stable region is compressed by a factor of ~4 compared to the input. This study shows that the simulation highlights new physical phenomena based on the interplay of various linear, non-linear and plasma effects that go beyond the experiment and is thus integral to achieving accurate designs of white light generation systems for optical applications. An adaptive error reduction algorithm tailor made for this simulation will also be presented in appendix.

## 1 Introduction

This paper describes a highly accurate and efficient simulation based on the novel methodology in [1] that incorporates a wide plethora of relevant linear and non-linear effects to model the white light generation (WLG) process that arises when an optical waveform propagates within a bulk material. Such as self-steepening, plasma effects, self-phase modulation (SPM), frequency dependent diffraction effects, and dispersive effects. Further to this, the simulation models any arbitrary input optical waveform in all three spatial dimensions and time ((3+1)D simulation) as it propagates within the material. A main contributor to the power of the simulation and the simulations' highest impact in terms of novelty to the community lies in its ability to accurately and efficiently model the complicated terms in the derivative generalized non-linear Schrödinger equation (GNLSE) that arise due to incorporating the self-steepening effect. This will extensively be explained in this paper. The new techniques used in the simulation to model the WLG processes makes it integral to obtaining a deeper understanding of WLG and to accurately design WLG systems for optical applications.

Understanding the WLG process is of utmost importance to the optics community since it is used for many optical applications such as seeding optical-parametric amplifiers (OPA) [2-5], two-dimensional spectroscopy [6] and non-linear compression [7]. White light generation is the process whereby the bandwidth of an optical pulse propagating in material undergoes substantial broadening. This is primarily due to the non-linear Self-phase modulation (SPM) effect that generates new frequencies, in a coherent manner through an intensity dependent additive phase. However, this is not the only relevant effect in the process; the influence of linear terms such as diffraction and dispersion and additional non-linear terms such as self-steepening and plasma generation plays a large role in WLG. SPM is directly related to the temporal derivative of the intensity and is prominent when the input pulse is in the range of femtosecond to picosecond timescales. For this reason, most WLG studies occur within this temporal range. Due to the peak gradients and intensities that exist at these timescales other limiting effects become relevant, an important one being the self-steepening and plasma effects.

Self-steepening of optical pulses arises because the intensity dependent refractive index influences the group velocities of the generated instantaneous frequencies under the amplitude envelope of the pulse. This causes an asymmetric temporal amplitude profile with a sharp cutoff and an elevated peak adjacent to the cutoff [8]. The nature of the self-steepening term within the GNLSE does not allow it to be simulated with conventional present day exponential Fourier split-step methods (EFSSM) [ REF balac \h \* MERGEFORMAT 9]. Because, a term that consists of a time-dependent coefficient in front of a time-derivative operator acting on the normalized optical amplitude emerges. In the past, Runge-Kutta and numerical difference methods were used instead [10-12]. However, they are proven to be less accurate and stable than EFSSM [13] and thus, comparatively converge to experimental results with much larger grid sizes. This means more computational resources must be employed to achieve the same accuracy which in turn limits the usefulness and power of the simulations. Therefore, to seek an updated EFSSM based method that can provide the stability and accuracy that these methods offer over Runge-Kutta and other methods to simulate WLG is of utmost importance. Understanding and designing effective WLG systems for optical applications or for fundamental physics can thus be pursued more accurately.

Given the above discussion, the simulation is based on the updated Strang EFSSM method derived in [1], which provides the full advantages and accuracy of an EFSSM that can be applied to a wider range of non-linear differential equations in all spatial dimensions and time. This includes the derivative GNLSE equation describing all above effects that go into WLG with the self-steepening terms. As a bonus, the method provides an intuitive viewpoint of the self-steepening process: The math corroborates the intuitive picture (discussed in section 4) which is not evident from the self-steepening term itself within the generalized NLSE.

The self-steepening effect must be modelled with plasma generation to fully capture the physics of WLG. The GNLSE used in the simulation is taken from [14] and includes all relevant effects with plasma absorption and scattering. Self-steepening enhances the temporal gradient due to the cutoff and consequently SPM. However, it ultimately causes the collapse of the pulse and material damage due to the growing peak intensity. As demonstrated in [14] the effect can be clamped and balanced by the creation of a plasma through multiphoton absorption; the pulse can propagate through longer crystals creating a larger frequency broadening without breaking apart or destroying the material. This clamping effect will be studied in this paper. Also, the creation of a near-temporal, spatial soliton (quasi-optical bullet) through the balancing of the various nonlinear, linear, plasma and self-steepening effects will be shown.

The application of the mathematical methodology to the GNLSE to obtain the numerical simulation will be described. Then to highlight the power of this simulation, results going beyond what was experimentally measured in a published example ytterbium aluminum garnet (YAG) system at 3.1 μm will be shown. The simulation indicates the creation of a near-optical light bullet in the experiment that was not discussed in the original paper, highlighting its importance to the community.

## 2 Defining the Equation and Explaining the WLG Process

This section will define the relevant equation that will be used for the simulations taken from [14]. This equation describes the WLG process in bulk material with the inclusion of all effects studied in this paper. The GNLSE is given as:

$$\frac{\partial u}{\partial \varsigma} = \frac{i}{4}\left(1 + \frac{i}{\omega_o \tau_p}\frac{\partial}{\partial \tau}\right)^{-1} \nabla_\perp^2 u - i\frac{L_{df}}{L_{ds}}\frac{\partial^2 u}{\partial \tau^2}$$
$$+ i\left(1 + \frac{i}{\omega_o \tau_p}\frac{\partial}{\partial \tau}\right)\left[\frac{L_{df}}{L_{nl}}|u|^2 u - \frac{L_{df}}{L_{pl}}\left(1 - \frac{i}{\omega_o \tau_c}\right)\rho u + i\frac{L_{df}}{L_{mp}}|u|^{2(m-1)} u\right] \quad (1)$$

$$\nabla_\perp^2 = \nabla_\chi^2 + \nabla_\psi^2$$

This is a non-unitary equation with plasma absorption terms. The above differential equation describes the evolution of the input envelope electric field normalized to the peak amplitude, represented as $u$. The first two terms are the linear terms of the equation. The rest are non-linear terms, involving functions of $u$. $\chi = \frac{x}{S_p}$, $\psi = \frac{y}{S_p}$ are unit-less coordinates of the transverse coordinates to the propagation coordinate direction. The equation is over normalized-to-input $e^{-1}$ values for the various dimensional coordinates. $S_p, \tau_p$ are the $e^{-1}$ values for the spatial extent and temporal duration of the input pulse intensity function. $\omega_o$ is the angular central frequency of the original pulse. The normalized time-coordinate is in a frame of reference travelling at the group velocity of the central frequency of the input pulse. The unit-less $z$ propagation coordinate is given as $\varsigma = z/L_{df}$, where $L_{df}$ is in meters and represents the diffraction length (the Rayleigh length for an input Gaussian). The $L$ constants represent various non-linear and linear lengths for physical processes and are listed in Appendix A.

$\rho$, the normalized plasma density term is a function of $u$. The optical radiation undergoes multi-photon absorption and avalanche ionization to produce plasma in the material. The plasma that is created is assumed to be static over the timescale of the pulse duration. The plasma density is defined by a linear first order non-homogenous differential equation:

$$\frac{\partial \rho}{\partial \tau} = \alpha \rho |u|^2 + |u|^{2m} \quad (2)$$

$m$ is a constant and is related to the order of photo-absorption.
Eq. (1) was derived under the slow-varying approximation, making it valid only in a frequency bandwidth equivalent to the central frequency of the input. The paraxial approximation is also used. It is verifiable that Eq. (1) is of the general form of the class of non-linear equations given in [1] and thus, the method in [1] can be used.

$u$ at the input is given as:

$$u_o = e^{-\left(\frac{x^2+y^2}{2S_p^2} + \frac{t^2}{2\tau_p^2}\right)}$$

Table 1, summarizes the physical meaning of each term on the right-hand side of the equation, as discussed in [14].

| Term | Physical Process |
| --- | --- |
| $\frac{i}{4}\left(1 + \frac{i}{\omega_o \tau_p}\frac{\partial}{\partial \tau}\right)^{-1} \nabla_\perp^2 u$ | Space-time Focusing (**linear term**): Diffractive term coefficient (function of temporal derivative). Accounts for the frequency specific diffraction of the optical radiation. |
| | Diffractive term (**linear term**): Accounts for spatial propagation in transverse dimensions along the propagation axis. |
| $-\mathrm{i}\dfrac{L_{df}}{L_{ds}}\dfrac{\partial^2 u}{\partial \tau^2}$ | Dispersion term (**linear term**): Approximation assumes constant group velocity dispersion (GVD) across generated spectral components. |
| $\mathrm{i}\left(1 + \dfrac{\mathrm{i}}{\omega_o \tau_p}\dfrac{\partial}{\partial \tau}\right)$ | Term including the self-steepening effect. |
| $\dfrac{L_{df}}{L_{nl}}\|u\|^2 u$ | **Non-linear term:** Describing Self-Phase Modulation (SPM) and Kerr- spatial lensing both due to the intensity dependent nature of the refractive index. |
| $-\dfrac{L_{df}}{L_{pl}}\left(1 - \dfrac{\mathrm{i}}{\omega_o \tau_c}\right)\rho u$ | **Non-linear term:** Plasma term describing plasma scattering and effects due to the refractive index variation of the plasma population. This is based on the Drude model. |
| $\mathrm{i}\dfrac{L_{df}}{L_{mp}}\|u\|^{2(m-1)} u$ | **Non-linear term:** Plasma absorption term describing the effect of multiphoton absorption. |

Table 1: Physical meaning of derived terms in Eq. (**1**). The non-linear terms are multiplied by the self-steepening coefficient, creating complicated non-linear terms that are solved by the mathematical methodology described in [**1**].

## 3 Defining the Split Step Operators and Split Step Method for the NLSE

In this section, the numerical algorithm used to solve Eq.(1) will be shown in a summarized fashion. The algorithm will be stated so that it could be used to recreate the simulation. However, the rigorous mathematical proofs and in-depth description that validates it will be omitted as it was already published in complete detail in [1]. Also, this algorithm scales with third order error with the propagation step size, i.e., $O(\varsigma^3)$. This is inline with traditional EFSSMs. When a Fourier transform is required within this algorithm, numerically it will be implemented using the fast Fourier transform (FFT) and inverse fast Fourier transform (IFFT). Eq.(1) differs from traditional NLSEs due to the following derivative terms, which are grouped together and labelled as the $\hat{C}$ operator:

$$\hat{C}(\chi,\psi,\tau)u(\chi,\psi,\tau) = \left(\frac{-1}{\omega_o \tau_p}\right)\left[\frac{L_{df}}{L_{nl}}|u|^2 \frac{\partial}{\partial \tau} - \frac{L_{df}}{L_{pl}}\left(1 - \frac{i}{\omega_o \tau_c}\right)\rho \frac{\partial}{\partial \tau} + i\frac{L_{df}}{L_{mp}}|u|^{2(m-1)}\frac{\partial}{\partial \tau}\right]u$$

These terms are obtained in the expansion of $\mathrm{i}\left(1 + \dfrac{\mathrm{i}}{\omega_o \tau_p}\dfrac{\partial}{\partial \tau}\right)\left[\dfrac{L_{df}}{L_{nl}}|u|^2 u - \dfrac{L_{df}}{L_{pl}}\left(1 - \dfrac{\mathrm{i}}{\omega_o \tau_c}\right)\rho u + \mathrm{i}\dfrac{L_{df}}{L_{mp}}|u|^{2(m-1)}u\right]$ in Eq.(1).

Traditional exponential Fourier split step methods cannot solve this term as its composed of a product of a distribution and derivative term that are over a mutual domain: The temporal domain. This differs from the strict grouping of derivatives with constant coefficients into one operator term (that happens to also be purely a linear term) and the remaining terms, composed only of distributions into another term (that is usually deemed the non-linear operator). This grouping is the case covered by traditional EFSSM methods and only cover a small subset of NLSE equations. Since this updated EFSSM method can cover these derivative terms, this algorithm uses for the first time an EFSSM to solve a derivative GNLSE (in 3+1 D). The extended EFSSM method is over three operators with the additional operator being these distributional-derivative terms. It will now be shown how the algorithm accommodates for this novel addition. To begin, the definition of the $\hat{C}$ operator is restated:

$$\hat{C}(\chi,\psi,\tau) = \left(\frac{-1}{\omega_o \tau_p}\right)\left[\frac{L_{df}}{L_{nl}}|u|^2 - \frac{L_{df}}{L_{pl}}\left(1-\frac{i}{\omega_o \tau_c}\right)\rho + i\frac{L_{df}}{L_{mp}}|u|^{2(m-1)}\right]\frac{\partial}{\partial \tau} \qquad (3)$$

Then, the operator is defined over four domains as such:

$$\overline{C(\chi,\psi,w',\tau)} = \left(\frac{-1}{\omega_o \tau_p}\right)\left[\frac{L_{df}}{L_{nl}}|u|^2 - \frac{L_{df}}{L_{pl}}\left(1-\frac{i}{\omega_o \tau_c}\right)\rho + i\frac{L_{df}}{L_{mp}}|u|^{2(m-1)}\right](-iw') \qquad (4)$$

Where $w' \in \mathbb{R}$. $w'$ is treated the same as the angular frequency variable. $w'$ is used to highlight that Eq. (3) is not Fourier transformed to its angular frequency representation to obtain the operator representation that will be used. In fact, only the derivative term is converted to its "angular frequency representation". The $u$ used in the $\hat{C}$ operator terms above is $u$ entering the iteration slice (the output of the previous slice). This is also true for the $\hat{B}$ operator shown below. This is in accordance with the mean-value approximation used in EFSSMs. $\rho$ is found by solving its ordinary differential equation using a Runge-Kutta $4^{th}$ order method (RK4), again with the values of $u$ entering the iteration slice. The RK4 method is sufficient for the non-homogenous first order ODE shown in Eq.(2).
The term is then applied to $u$ as:

$$u_{k+1} = \text{IFFT}_{w' \to \tau}\left[e^{\frac{1}{2}\overline{C(\chi,\psi,w',\tau)}\Delta\varsigma} u_k(\chi,\psi,w')\right] \qquad (5)$$

$u_k$ is $u$ outputted from the previous operator step. $u_{k+1}$ is $u$ after the application of the $\hat{C}$ operator. This will be used as the input to another operator in the iteration (more details of the global iteration will be shown at the end of the section). The operations in Eq. (10) explicitly mean:

1. $u$ (i.e., $u_k$) is inputted in its spatial-frequency domain representation[1] because its frequency representation is the same as its representation in the $w'$ domain, (i.e., $u(w',\chi,\psi) = u(w,\chi,\psi)$).
2. The exponent is multiplied into $u(w',\chi,\psi)$ across $w'$ at a value of $\chi,\psi,\tau$.
3. The value of the inverse Fourier transform on the $w'$ domain of the new function (created in 2.) only at the value of $\tau$ used in 2. is taken. This is the value of the updated $u$ at $\chi,\psi,\tau$ coming out of the exponential $\hat{C}$ operator step.
4. The process is repeated for all $\chi,\psi,\tau$. At the input of this step $u(w,\chi,\psi)$ is sent and after this step an updated $u(\tau,\chi,\psi)$ is found[2].

The other operators notably the traditional linear operator (over derivatives with constant coefficients) and nonlinear operator (terms that are composed of distributions without derivatives) can be solved by the more traditional techniques used in EFSSMs.

The linear operator is labelled as:

---

[1] $\text{IFFT}_{k_\chi,k_\psi,w \to \chi,\psi,w}$ refers to a two-dimensional inverse Fourier transform only over the momentum coordinates.
[2] A 3-D function (representing $u$) is inputted into this exponential $\hat{C}$ operator step. The exponential $\hat{C}$ operator is a 4-D function. During the application of the step, the new function found in 2. Is 4-D. After step 3 is applied over all $\chi,\psi,\tau$ the function is a 3-D function.

$$\hat{A}(\chi,\psi,\tau) = \frac{i}{4}\left(1 + \frac{i}{\omega_o \tau_p}\frac{\partial}{\partial \tau}\right)^{-1} \nabla_\perp^2 - i\frac{L_{df}}{L_{ds}}\frac{\partial^2}{\partial \tau^2} \quad (6)$$

And is treated in the Fourier domains, so that the derivatives have functional representations that can be solved exactly:

$$\hat{A}(k_\chi, k_\psi, w) = -\frac{i}{4}\left(1 + \frac{1}{\omega_o \tau_p}w\right)^{-1}(k_\chi^2 + k_\psi^2) + i\frac{L_{df}}{L_{ds}}w^2 \quad (7)$$

$w$ is the angular frequency of $\tau$, $k_\chi, k_\psi$ are the angular frequencies of $\chi, \psi$. The region of validity for the series convergence in the inverse space must be considered, see Appendix C. The operator series expansion in the frequency domain (obtained from the binomial expansion w.r.t the time derivative of the inverse coefficient and then converted into the frequency domain) converges to Eq. (7) within the bandwidth of the slow varying approximation used for Eq. (1). $w$ is related to the angular frequency ($\omega$) of $t$ (proper time), for a pulse centered at $\omega_o$ as:

$$w = \tau_p(\omega - \omega_o) \quad (8)$$

Where, $\omega_o$ is the central angular frequency of the input pulse. $w$ is referred to as the "reduced frequency" from here on.

The remaining coefficient terms in Eq. (1) are functions of $u, \rho$. This can be placed into the traditional nonlinear operator of EFSSMs. The nonlinear operator is labelled as:

$$\hat{B}(\chi,\psi,\tau) = i\left[\frac{L_{df}}{L_{nl}}|u|^2 - \frac{L_{df}}{L_{pl}}\left(1 - \frac{i}{\omega_o \tau_c}\right)\rho + i\frac{L_{df}}{L_{mp}}|u|^{2(m-1)}\right]$$
$$+ \left(\frac{-1}{\omega_o \tau_p}\right)\left[\frac{L_{df}}{L_{nl}}\frac{\partial}{\partial \tau}|u|^2 - \frac{L_{df}}{L_{pl}}\left(1 - \frac{i}{\omega_o \tau_c}\right)\frac{\partial}{\partial \tau}\rho + i\frac{L_{df}}{L_{mp}}\frac{\partial}{\partial \tau}|u|^{2(m-1)}\right] \quad (9)$$

In contrast to $\hat{A}$, it would be of no benefit to consider $\hat{B}$ in any inverse space and it is considered in the original $\chi, \psi, \tau$ space.

The general iteration scheme governing the application of these operators to $u$, starts with dividing the crystal into propagation slices that are solved iteratively. Per propagation slice the operators are applied in an iterative fashion. However, the operators do not commute and a Strang symmetrisation scheme must be implemented to reduce this error [9]. The symmetrisation relies on how fast operators vary with the propagation coordinate. For example, the fast-varying dispersion and diffractive effects must be sampled more. So, the $\hat{A}$ operator must be applied more on $u$ per propagation slice. The $\hat{C}$ operator describes the salient physics of the self-steepening effect (as discussed in section 4). Self-steepening must be sampled more as it can cause fast rising peak intensities if not properly balanced with dispersion, diffraction, plasma generation and SPM. Therefore, the $\hat{C}$ operator will be sampled more in the propagation slice. The $\hat{B}$ operator contains effects more dependent on the pulse envelope and are more slowly varying. Consequently, it is sampled once. Table 2 summarizes the physical interpretation of each operator. However, experimentation with the ordering is done to achieve better simulation results. These considerations are factored into the symmetrisation scheme used in this updated EFSSM to obtain the overall series of operations per propagation iteration slice:

$$Z = IFFT_{k_\chi,k_\psi,w \to \chi,\psi,\tau} e^{\frac{1}{4}\hat{A}(k_\chi,k_\psi,w)\Delta\varsigma} FFT_{\chi,\psi,\tau \to k_\chi,k_\psi,w}$$

$$IFFT_{w' \to \tau} e^{\frac{1}{2}\overline{C(\chi,\psi,w',\tau)}\Delta\varsigma} IFFT_{k_\chi,k_\psi,w \to \chi,\psi,w} e^{\frac{1}{4}\hat{A}(k_\chi,k_\psi,w)\Delta\varsigma} FFT_{\chi,\psi,\tau \to k_\chi,k_\psi,w} \qquad (10)$$

Eq. (10) yields:

$$u(\chi,\psi,\tau,\varsigma') = Z e^{\hat{B}\Delta\varsigma} Z u(\chi,\psi,\tau,\varsigma' - \Delta\varsigma) \qquad (11)$$

Eq. (11) is iteratively implemented over all steps in $\varsigma$.

| A | B | C |
| --- | --- | --- |
| Spatio-Temporal focusing, dispersion, diffraction | SPM, Kerr Lensing, plasma effects on refractive index, plasma scattering, plasma absorption, intensity envelope and plasma envelope contribution to self-steepening | Remaining Self-steepening contributions: Derivative of amplitude electric field |

Table 2: Physical interpretation of each operator used in the specific WLG bulk problem.

An adaptive step-size algorithm is shown in Appendix B. This algorithm updates step-sizes such that aliasing does not occur and that sampling step-sizes are always below the changing Nyquist criteria as the pulse acquires additional bandwidth and spreads temporally and spatially. The step-size algorithm accounts for operator specific sampling conditions and reveals more subtle physics that each operator accounts for. However, the algorithm has not been implemented fully in the simulation presented in this paper.

## 4 Inclusion of the Physics of Self-Steepening within the B and C operators

The novelty of the method is not only in the mathematics of solving the distributional-derivative terms of GNLSEs such as Eq. (1), but also doing so in a manner that can directly give the physical effects covered by these terms. This is a powerful consequence of this method as one can directly see the contributions of the physical effect in the simulation and can manipulate it. In this case, the physical effect covered by the distributional-derivative terms in Eq. (1) is the self-steepening effect. This effect will briefly be described below, and how the method directly gives the physically intuitive picture of the effect will be shown.

Self-Steepening imposes a group velocity delay at different intensity points along the pulse. Because, of the intensity dependent refractive index term that causes an intensity dependent group velocity (GV) for the instantaneous frequency at an intensity point of the pulse [15]. This effect contributes to a shift of the peak intensity and an asymmetric steepening of the optical pulse. This is graphically shown in Figure 1a.

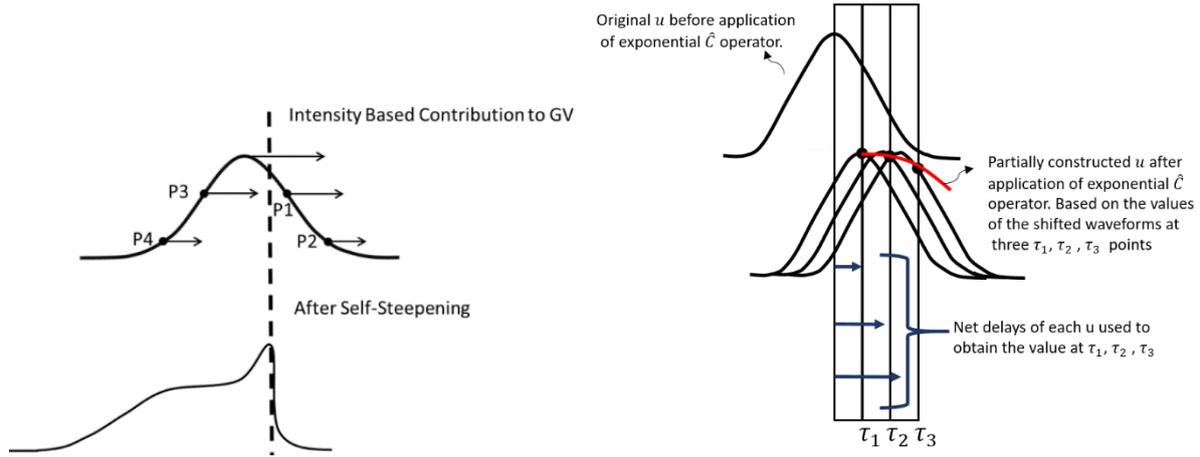

Figure 1: a) Self-Steepening imposes a group velocity delay at different intensity points along the pulse because of the intensity dependent refractive index term. For example, P1 has a higher velocity and heads towards P2 at the front of the pulse. P4 has a lower velocity than P3 and thus the net delay between these two points increases. The peak also shifts at a maximal velocity. This gives a steepened edge on the front of the pulse and a shifted peak towards the front, creating an asymmetric pulse profile. The amplitude is then corrected such that energy is conserved. b) An example that demonstrates the $\tau$ domain shifting effect of the exponential $\hat{C}$ operator using three $\tau$ domain values.

For the imaginary arguments of the exponential $\hat{C}$ operator, at a given $\tau$ value, it can be visualized that the effective operation is to take the input $u$ and shift it by $f(\tau, \chi, \psi)$ in the $\tau$ domain[3]. Then, the value of this shifted $u$ at the given $\tau$ is taken for the new value of $u$ after the application of the operator at this $\tau$. $f(\tau, \chi, \psi)$ is the coefficient of the exponential $\hat{C}$ operator of $(-iw')$. This coefficient, represents the group velocity delay because of the intensity dependent refractive index (i.e., because of an intensity dependent group velocity). This is demonstrated in Figure 1b and will create an effect shown in Figure 1a. However, the amplitude must additionally be increased or decreased such that energy is still conserved. For example, as two points on the amplitude envelope move closer together, the amplitude of the second point must increase to compensate for the shortened distance between the two points. This is factored into the real envelope-derivative arguments of the exponential $\hat{B}$ operator. The delay features of the exponential $\hat{C}$ operator are in line with the physically intuitive picture of the self-steepening effect. It demonstrates in a direct way that an intensity dependent group velocity delay brings values of $u$ closer or further apart in the $\tau$ domain.

The phase contributions of the exponential $\hat{C}$ operator in the $\tau$ domain can yield additional insights to what was described in the paragraph above. There are additional subtle effects on the instantaneous frequency distribution of $u$ after the application of this operator. This will now be shown. The exponential $\hat{C}$ operator reorganizes the phase information from the phase function of the input $u$ ($\varphi_o(\chi, \psi, \tau)$) as follows:

$$\varphi_C = \varphi_o(\chi, \psi, \tau - f(\tau, \chi, \psi)d\varsigma) \tag{12}$$

Labelling,

---

[3] This is due to the shift identity of Fourier transforms

$$G = \tau - f(\tau, \chi, \psi)d\varsigma \qquad (13)$$

The phase derivative or instantaneous frequency is given as:

$$\frac{\partial \varphi_C}{\partial \tau} = \frac{\partial \varphi_O(\chi, \psi, G)}{\partial G}\left[1 - \frac{\partial f(\tau \chi, \psi)}{\partial \tau} d\varsigma\right] \qquad (14)$$

If the multiplication in Eq. (30) is expanded, the first term equates to shifting the existing instantaneous frequency distribution by the shift procedure described for the exponential $\hat{C}$ operator above. However, the shift procedure does not simply shift the existing instantaneous frequency distribution, since it shifts values of $u$ and not only its instantaneous frequency. For example, when it brings two values of $u$ or further together in $\tau$ it also brings the corresponding phase values of these points closer or further together. This changes the instantaneous frequency between these two values. The second term accounts for this additional adjustment of the instantaneous phase. Physically, it can be viewed as SPM that occurs due to intensity dependent GVs that stretches or compresses the envelope and thus the phase function that it carries (i.e. like stretching and compressing a spring: The spacing period between turns in the spring will change along with the spring being shorter or longer.). In conclusion to this section, the physics of the self-steepening effect are included in the exponential $\hat{C}$ operator in a physically intuitive manner where the math can directly be related to the physics. The algorithm not only solves the new distributional-derivative terms in Eq. (1) but also does this so that the physical effect can be seen directly from the math. The physical picture is not lost because of complicated mathematical acrobatics. Also, the effect can still be separated out in the math and not obscured by being mixed with others. It also makes the effect more apparent than what can be deduced from the original term in the equation.

## 5 Numerical Results from the WLG Simulation

The overall goal of this results section is to cover two major points:

1) *The simulation is valid:* The YAG system from [16] is well described in terms of all input and material parameters. As well, the data was simulated using another technique in that paper. Therefore, to validate this simulation a comparison between it and the results from [16] will be presented.
2) *The simulation is useful and new physics can be predicted:* After validation of the simulation, it is then important to show that the simulation can predict and yield more insights and go further than what was described in [16]. A complete study of the spatial profile, the temporal characteristics and the contributions of various effects including self-steepening and plasma will be discussed. New and interesting results will be highlighted. An interesting regime is found from the simulation where a near-temporal and spatial soliton exists, i.e., a "light bullet".

### 5.1 Verification of Simulation

The experimentally found spatially integrated spectral density from [16] was obtained with the most important parameters listed in Table 3. There was a slight deviation in the simulation values used for the peak intensity and the spatial spot size. Because, the experimental measurement for the spectral energy density plot was done at a smaller value for the spatial spot size and pulse energy. The exact value was not listed in the paper. The complete set of parameters including relevant material constants are found in Appendix A. Appendix B lists all the simulation-specific values such as sampling intervals, conditions, etc. Material values were taken directly from [16] without change.

| Parameter | Experimental Value | Simulation Value |
|---|---|---|
| Spatial $e^{-2}$ of the intensity | 50 µm | 42.4 µm |
| FWHM temporal intensity duration | 85 fs | 85 fs |
| Peak power | 76 MW | 40 MW |
| YAG crystal length | 2 mm | 2 mm |
| Input central wavelength | 3.1 µm | 3.1 µm |
| Spatial profile | Radial Gaussian | Radial Gaussian |

A pertinent note: When discussing the temporal properties of the optical pulse, the time coordinate is in a frame of reference travelling at the group velocity corresponding to 3.1µm.

Table 3: A cross-section of relevant parameters listed in [16]. Simulation parameters used in obtaining the fit for the spectral energy density shown in Fig. 3 are also shown.

The comparison between this simulation technique, the experimental data and the simulation from [16] is shown in Figure 2. The bandwidth range exceeds the slow-varying approximation used in the derivation of the white-light equation shown in section 2. However, a good match was still found. The range of the simulation was set at 1700 nm, due to the zero dispersion point in YAG by zero-padding for wavelengths lower than 1700 nm and higher than the absorption edge of YAG (at 4700nm). This range limitation is due to the assumption of Eq. 1 that the GVD is unchanging through the spectral window. Other sources of error include the fixed absorption order and all that is discussed in Appendix C. The radially symmetric Gaussian input that was assumed in the experiment may not have been the case. Figure 2 indicates less than a factor of 5 deviations between the experimental data and the simulation technique herein described. As well, the simulation presented in [16] deviates considerably from this simulation and the experimental data to over an order of magnitude. This is especially visible on the blue side of the spectrum. Thus, this simulation method outperforms previous WLG simulation techniques, such as the one used in [16]. The results presented were obtained at numerical convergence indicated in Table 4 of Appendix B. The self-steepening effect manifests itself in the blue pedestal of both the experimental and simulation curves of Figure 2.

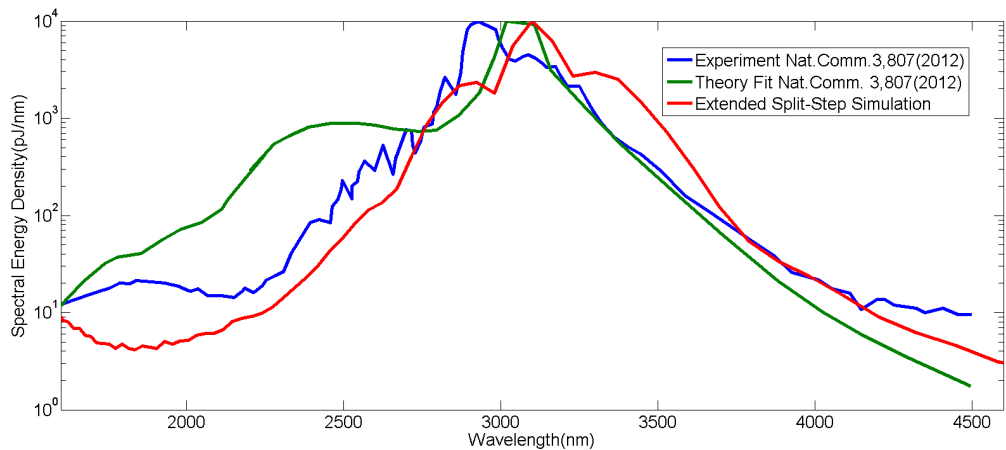

Figure 2: Comparison of theory fit (shown in green) and experimental data (shown in blue) [4] from [16] and the simulation (shown in red). Simulation fit is in excellent agreement to the experimental data (≤factor of 5 everywhere). The theory fit in [16] diverges considerably on the blue side of the spectrum.

---

[4] Data was acquired using the Grabit! App.

From Figure 2 it can be deduced that the simulation presented in this paper is validated at least for the experimental parameters used in [16] and the bandwidth presented in the figure. These parameters will be used for the rest of this section. It should be noted that both the simulation and experimental bandwidth shown in [16] extended into the visible range and was substantially broader than 1700nm. The simulation here is limited due to the changing sign of the GVD at the zero-dispersion point that would severely violate the assumption used in Eq.(1), that the GVD is relatively static within the bandwidth range.

*For the continuation of this section, all figures in the temporal domain are in the temporal frame travelling at the group velocity of the inputted central frequency of the pulse.*

## 5.2 An Interesting Story: A Near Temporal Spatial Soliton (Light Bullet)

Since SPM and self-steepening are intensity dependent, the temporal effects are coupled with the intensity modifying SF and plasma effects. Thus, the nonlinear temporal and spatial effects interact through the common intensity term. This interaction is more interesting and relatively not as well-known as the traditional SF plasma focusing-defocusing cycles (plasma filamentation) that create a self-guiding filament [8]. Where, the pulse dynamically maintains its spatio-temporal features in a constrained range. As will now be described, in the context of the experiment in [16], this interaction can also produce focusing-defocusing cycles (like plasma filamentation) so that the pulse is self-guiding or other interesting features. These effects are of interest within the optics community to obtain stable spatial-temporal solitons [17-24], i.e., optical bullets. Optical bullets preserve their spatio-temporal pulse properties over large distances (they are propagation invariant [22]) by self-guiding processes and have been the focus of many studies that seek to experimentally or theoretically obtain them [21- 24]. The demonstration of a near optical bullet in the experiment of [16] is a new result that has not been discussed before and is interesting given the importance in achieving optical bullets to the community. Thus, its demonstration justifies the use of this full (3+1)D simulation as it accomplishes item 2) in the list at the beginning of section 5. Section 5.2 will start with discussing the spatial fluence distribution of the optical pulse, then in 5.2.2, pulse characteristics leading to the formation of the near-optical bullet will be discussed. Finally, section 5.2.3 will end with discussing the near-optical bullet present in the experiment.

## 5.2.1 Radial Symmetry Numerical Test and Characteristics of the Spatial Fluence as the Optical Pulse Propagates

To start with, the simulation does not assume any spatial symmetry such as radial symmetry so it can model any arbitrary spatial distribution of inputted optical pulses. As shown in Appendix B the transverse spatial array is a two-dimensional rectangular grid. As well, it can be shown that all terms in Eq. (1) preserve radial symmetry. Therefore, a good numerical test of the rectangular grid size and spacing is to simulate a radially symmetric input spatial distribution (as is done here to match the example system of [16]). To see if radial symmetry is conserved throughout the simulation. Figure 3b shows the end spatial fluence plotted across the transverse dimensions and is a good example that shows that at numerical convergence, the radial symmetry is preserved.

The focusing-defocusing cycles of the optical fluence is shown in Figure 3a, for the same parameter set as used in Figure 2. This was not shown in [16] and to experimentally get access to this information would rely on complex interferometric measurements [20], highlighting the usefulness of a full spatial and temporal simulation. From Figure 3a, there are two focusing-defocusing cycles, after which the pulse appears to be collimated at a smaller radius. Not only is the fluence collimated but it will be seen that there is a near-temporal soliton effect as well.

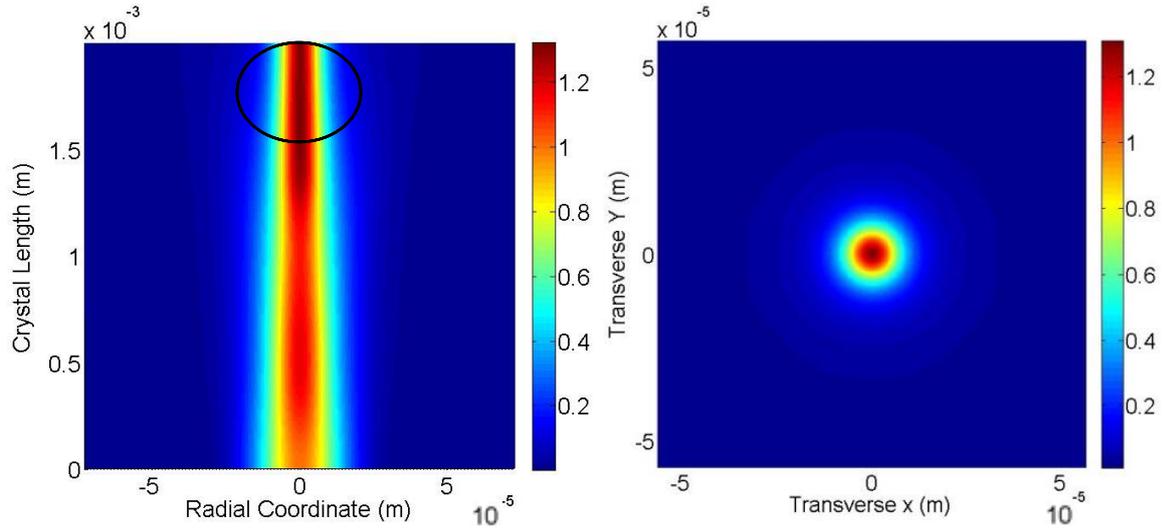

Figure 3: a) Pulse fluence normalized to peak input fluence plotted for the radially symmetric input in the system described in [16]. Circle indicates collimation point where plasma and SF effects balance. b) Normalized spatial fluence distribution at the end of the crystal in all transverse dimensions. The above demonstrates that radial symmetry is preserved for a radially symmetric input, which is a good spatial grid size check for the simulation.

### 5.2.2 The Nonlinear Shaping Dynamics to a Self-Steepened Pulse and its Effect on the Plasma Filamentation.

The first cycle, shown in Figure 3a from ~0.2 to 0.7 mm, is weekly interacting with the plasma and can be viewed as a plasma independent cycle. Because, in the matching region in Figure 4a, the total pulse energy stays constant; meaning no plasma absorption. Thus, this mimics the focusing-defocusing cycles of plasma filamentation without the plasma (a result which is not discussed in literature). Within this first cycle, the pulse is non-linearly shaped in intensity such that temporal compression (due to material parameters, SPM shifts frequencies in the opposite sense of dispersion) and rising peak intensities exist along the optical center. As with plasma effects, self-steepening plays a minimal role in the temporal intensity profile. This is shown in Figure 5a, where the characteristic features of self-steepening are subdued. At the end of this region, the temporal intensity profile along the optical center is compressed to a $\text{sech}^2$ profile. From ~0.7mm to 1.2mm the temporal intensity profile along the optical center stays as a $\text{sech}^2$ but with rising peak intensity (shown in Figure 4b ). Still there is no significant plasma absorption or self-steepening. From 1.2-1.5 mm the high gradients and peak intensities of the $\text{sech}^2$ temporal intensity profile, initiates self-steepening. Figure 5b shows the optical pulse being shaped into the self-steepened pulse within this propagation region. Due to the high peak intensities that occur during this shaping, plasma generation starts to take effect as seen in the levelling off peak intensity in Figure 4b and the gradually decreasing region of the exponential function in Figure 4a. Figure 3a indicates that this corresponds to the focusing region that directly feeds into to the collimated focused region (circled in the figure). Thus, when self-steepening is initiated, plasma generation is activated.

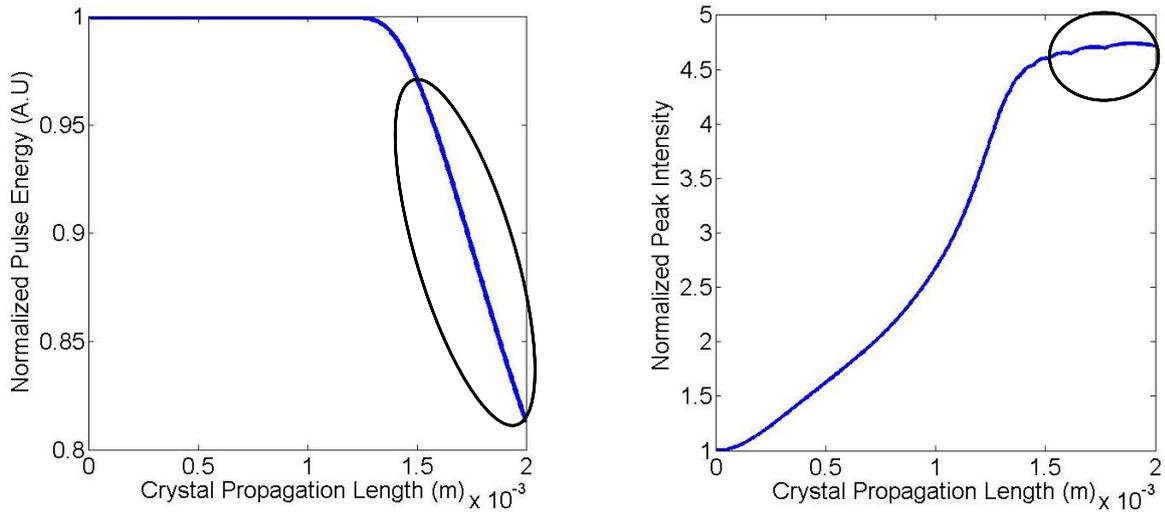

Figure 4: a) Normalized-to-input pulse energy as it propagates in the crystal. The onset of plasma generation and the start of the plasma filamentation occur roughly at 1.3 mm. The circled region corresponds to the circled region of Figure 3a and shows heavy plasma absorption in the non-linearly "collimated" region. The above shows that roughly 20% of the input pulse energy is lost due to multiphoton plasma absorption effects. b) Normalized peak intensity along optical center. Saturated region (circled) corresponds to circled region in a) and Figure 3a.

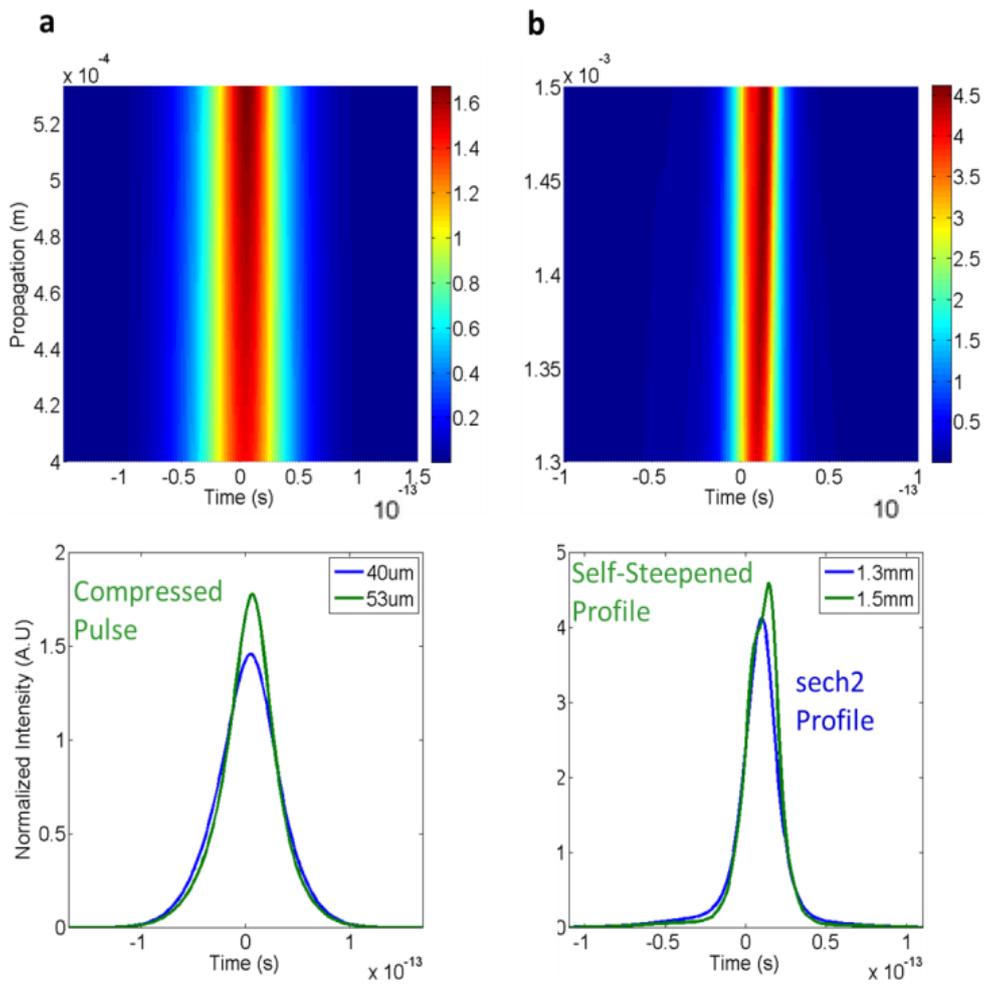

Figure 5: a) Normalized to input peak intensity of temporal profile of optical center in the region of the first focusing peak in the fluence. Non-linear pulse compression begins to take place and a slight peak shift due to the self-steepening effect is visible. b) Normalized to peak intensity of the temporal profile along the optical center 1.3-1.5 mm within the crystal. Self-steepening and plasma generation begins in this region. The non-linearly compressed pulse is furthery compressed until self-steepening becomes significant due to the rising peak intensities and gradients, which then dominates shaping of the pulse.

The characteristic self-steepening effect manifests itself roughly around 1.5 mm within the crystal. This is seen in Figure 5b and Figure 6, which plots the
amplitude, intensity and phase temporal profiles from 1.8 mm onwards. The symptomatology of the self-steepening effect; i.e., The intensity peak shift from the input peak (at time zero) to the back of the pulse[6], the cutoff and compression on one side of the pulse and the stretching on the other side is clearly visible in the amplitude/intensity plot of Figure 6a. Furthermore, the effect is also clearly seen in the phase compression around the cut-off side and stretching on the other side shown in the phase plot of Figure 6b. Also, in accordance to what is expected for the material values of YAG, frequencies on the blue side accumulate on the cut-off side of the pulse. Because of the self-steepened pulse, the peak intensity is high enough to generate significant multi-photon ionization as can be seen by the linear portion of the exponential decay function in Figure 4a. This is where significant plasma absorption takes place and a filamentation forms.

It can thus be concluded that in this experiment self-steepening ultimately generates the peak intensities that trigger the plasma filamentation. When SF is dominant but not self-steepening (0-1.2 mm in the crystal) negligible plasma generation happens as seen in Figure 4.

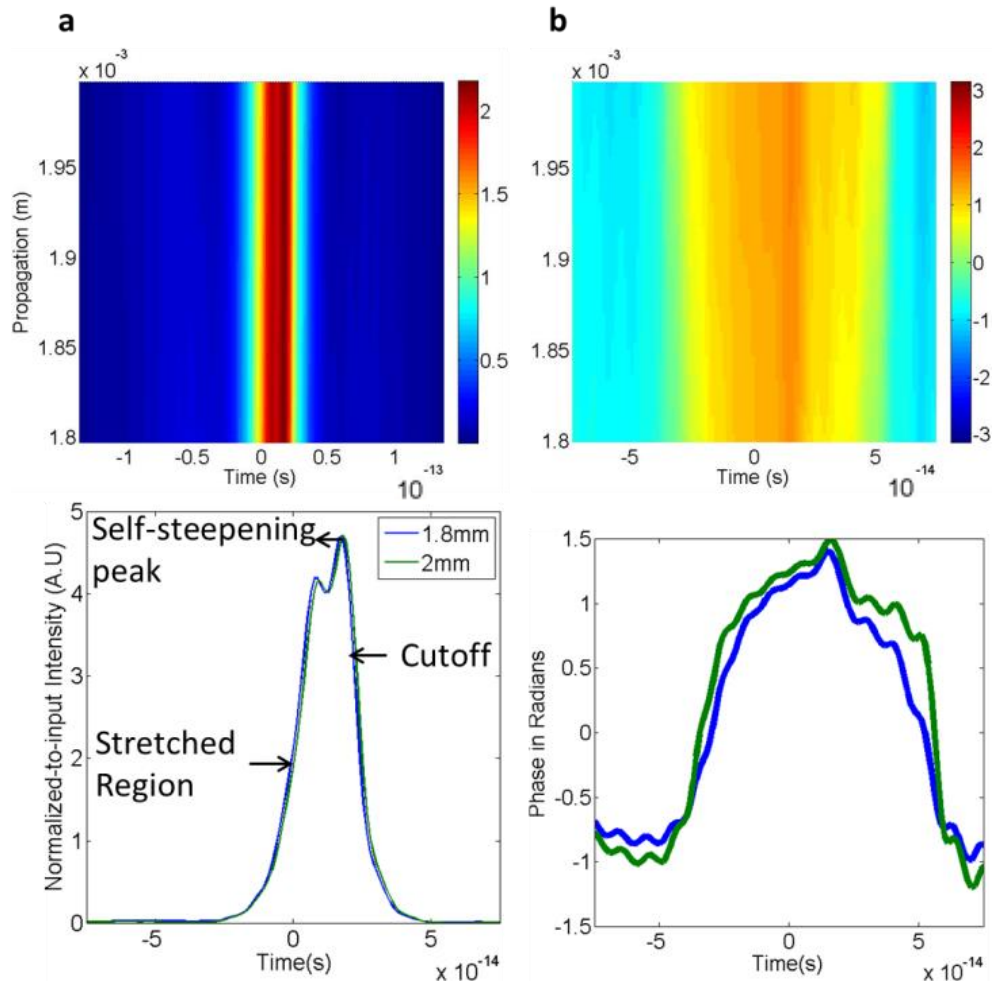

Figure 6: a) Normalized temporal **amplitude** profile (within relevant time range) along the optical center, plotted against crystal propagation distance from 1.8 to 2 mm. **Intensity** (normalized to input peak intensity) profiles at the beginning and end of the propagation region shown **below**. The intensity plots show the tell tale marking of the self-steepening effect. b)

---

[6] This is the case when the material has a positive $n_2$ and negative $GVD$ within the range of frequencies considered, which is the case for YAG.

The phase profile in radians as a function of the same propagation distance as a), plotted in the time region where the intensity is non-negligible. Profiles at the beginning (blue) and end (green) are shown below. The amplitude, intensity and phase profiles are near non-varying as a function of propagation distance especially where self-steepening dominates.

### 5.2.3 The Formation and Maintenance of the Near Optical Bullet

The shaping to the self-steepened profile is now complete and for the remainder of the crystal the self-steepened profile will be maintained. As seen from Figure 6a,b, from 1.5-2 mm in the crystal the temporal intensity and phase profile of the optical center does not change significantly as it propagates, i.e., becomes clamped. Especially where the self-steepening effect dominates (near the location of the main intensity peak). This also corresponds to the saturation region of the peak intensity shown in Figure 4b. The self-steepening profile is clamped by multi-photon absorption producing the plasma. Therefore, a near temporal "soliton" exists from 1.5 to 2 mm in the crystal. In contrast to the simple textbook temporal soliton case of the 1-D NLSE, where SPM and dispersion counter and balance each other [8], the net frequency generation and temporal broadening/compression of SPM and dispersion is now balanced with the amplitude reshaping and frequency generation of the self-steepening term and the heightened plasma absorption. However, due to the plasma absorption there should be an overall decrease in intensity across the profile. The reason why this is not the case is because of the spatial effects. This is where the self-focusing effect comes into play. It is the last element of the balancing act moving energy from the wings of the pulse into the optical center. Figure 3a shows that there is a slight focusing effect in the wings. The inbound energy due to SF matches with the plasma scattering and absorption effect that removes energy from the optical center. Self-focusing also helps to maintain a clamped fluence by balancing with the plasma defocusing and absorption effect within a range of ~13.16μm around the optical center, within this propagation region. This produces a spatial "non-linear" collimation effect and explains the fluence collimation in the circled region of Figure 3a. The collimated spatial FWHM extent is ~1.88 times smaller than the initial input, as shown in Figure 7a. It can then be concluded that within ~13.16μm around the optical center the pulse is non-varying spatially and temporally as it propagates within this crystal region creating a near-optical bullet. Also the bullet's peak fluence is the highest in the crystal at 1.3 times the original peak fluence.

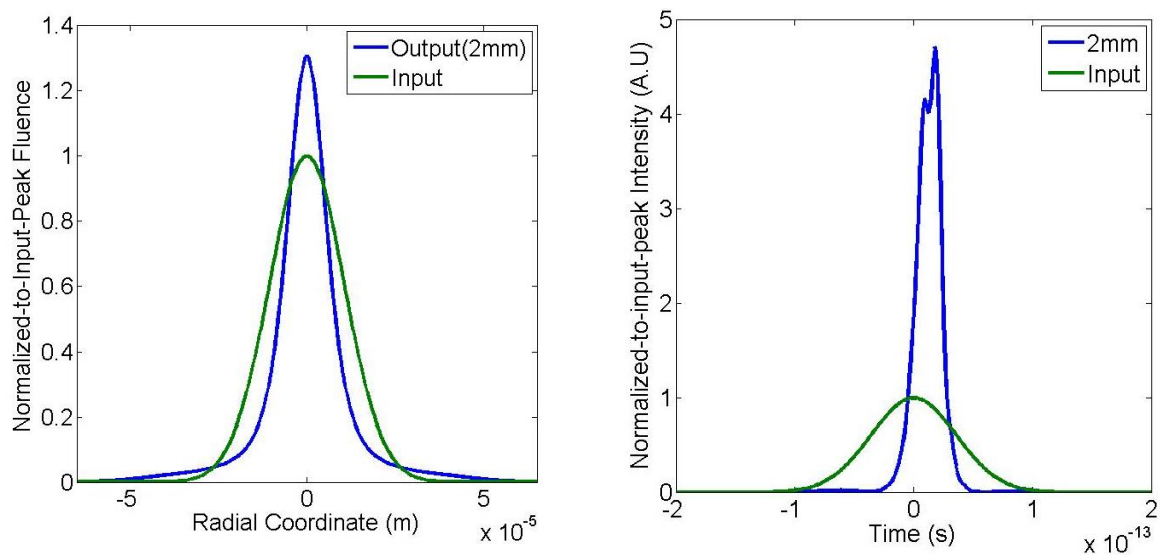

Figure 7: a) A comparison between the normalized Gaussian input fluence and the output fluence. The output FWHM is ~1.88 times smaller at the output and is a hyperbolic secant squared profile. b) A comparison of the input and output of the normalized temporal intensity function along the optical center. The output has a net FWHM compression factor of ~3.88. The self-steepening effect shifts the peak 17.9fs to the back of the original input pulse.

As a bonus, much like the smaller spatial extent of the bullet, Figure 7b shows that the temporal intensity along the optical center is at a compressed FWHM duration that is approximately a factor of

~3.88 from the original input pulse. The peak intensity at the output is ~4.72 times larger than the input peak intensity. There was also a net self-steepening peak shift of 17.9fs towards the back of the original input pulse. This spatial and temporal compression is expected: For this spatio-temporal collimation to occur high peak intensities, temporal and spatial gradients must be present to initiate the necessary nonlinear effects. From the above observations, the propagation region from 1.5 to 2 mm in the crystal is of utmost interest due to the near optical bullet that is present.

This spatially collimated fluence and near temporal soliton (at least along the center of the optical pulse) can partially explain the shot to shot stability of the WLG reported in [16]. The balancing of the non-linear and linear effects along a comparatively large region of the crystal (approx. 25% of the crystal) indicates that there is a large stable point in parameter space of the non-linear differential equations describing the system. Therefore, the system is more tolerant to perturbations in the optical waveform at the entrance of the crystal.

The above proves the importance of the new simulation by satisfying the second point in the list of the introduction of section 5. The simulation can highlight new physics that can go beyond the original experiment and can model intricate effects and their interplay, such as the self-steepening plasma interplay of the example YAG system. A highly interesting physical case emerges from the simulations, one that is sought after by the community: The near perfect soliton dynamic: where a spatial (shown in terms of fluence) and temporal soliton can exist.

# 6 Conclusions and Extensions

A novel and fast three dimensional + time simulation technique based on an updated symmetric exponential Fourier split-step method has been used to model WLG in bulk material including the self-steepening term. This algorithm can be applied to arbitrary dimensional derivative GNLSEs such as the equation used in this paper to model WLG with self-steepening. It was also shown that this algorithm was physically intuitive and directly corresponded to the physical effects being modelled (it is easy to see the physical effect directly from the math, without many steps).

The simulation and its results were compared to published experimental data, given in [16], on an example 2 mm bulk YAG crystal with an ultrafast input optical pulse centered at 3.1 µm. Deviation between the simulation and the experiment was less than a factor of 5. This deviation was substantially less than that of the simulation that was originally used to describe the experiment, also presented in [16]. Using the simulation, a more rigorous picture of the experiment highlighting interesting and new physics was explored, especially high-lighting the effects of the self-steepening as the optical profile propagates. It was shown that the optical waveform is shaped into a quasi-light bullet near the end region of the crystal used in the experiment. A complete view of the important aspects of WLG can be achieved with this simulation as was shown extensively in the results section. The notable results of the simulation, presented in this paper, demonstrates the importance of this simulation and the updated technique it relies on for nonlinear optics problems. For example, with these simulations systems that can control wave-breaking and material damage can be built for WLG applications. Stable points can be explored for various parameter sets.

This paper lays the foundational description and proof of convergence to experimental results of the simulation. Future studies will validate the simulation across multiple experimental conditions, such as different input central wavelengths, pulse energies, waveforms and materials. As well, an adaptive algorithm to avoid under-sampling was derived in this paper that is tailor-made for the new methodology. This is shown in Appendix B. This algorithm will be extensively tested numerically. Additional extensions, supported by the mathematical methodology, include adding Raman terms and to account for the polarization of light. The simulation's accuracy is limited by the accuracy of the GNLSE equations that the numerical technique models. The only constraints of the numerical technique are the Nyquist criterion for Fourier based methods, the mean field approximation in relation to the propagation coordinate and the commutation error of the operators.

## Competing Interests

The author does not have any competing interests.

## Acknowledgements

The author would like to thank Axel Ruehl and Aradhana Choudhuri for pointing the author to reference [14]. The author would like to thank Aradhana Choudhuri for her help at the beginning of coding a draft of part of the method on the MATLAB platform. The author would like to thank Prof. Dr. R.J. Dwayne Miller, for the use of the research group's computational resources and for funding.

## Funding

Funding for this work was supported through the Max Planck Institute for the Structure and Dynamics of Matter.

# Appendix A: Explanation of Constants in the GNLSE

| Symbol | Description | Value |
| --- | --- | --- |
| $c$ | Speed of light | $2.997(10^8)\frac{m}{s}$ |
| $n_2$ | Nonlinear refractive index | $7(10^{-20})$ |
| $\omega$ | Central angular Frequency | $6.08(10^{14})s^{-1}$ |
| $\beta_2$ | GVD | $-4.08(10^{-25})\frac{s^2}{m}$ |
| $k_o$ | In material Wavenumber corresponding to: $\omega$ | $3.61(10^6)m^{-1}$ |
| $m$ | Order of photon absorption | 17 |
| $\beta^m$ | mth photon absorption coefficient | $7.63(10^{-266})m^{31}W^{-16}$ |
| $\sigma$ | Inverse Bremsstrahlung cross section | $2.6(10^{-24})m^2$ |
| $\tau_c$ | Electron collision time | $3(10^{-14})s$ |
| $E_g$ | Ionizing Potential Energy | $6.5eV$ |
| $A_o$ | Peak amplitude of the envelope electric field | $2.2\frac{V}{m}$ |

Appendix A Table 1: A list of the symbols and values of pertinent quantities used in the simulation for the example YAG system.

| Symbol | Name | Equation |
|---|---|---|
| $L_{nl}$ | Nonlinear length | $\dfrac{c}{\omega n_2 I_o}$ |
| $L_{ds}$ | Dispersion Length | $\dfrac{\tau_p^2}{\beta_2}$ |
| $L_{df}$ | Diffraction Length | $\dfrac{k_o S_p^2}{2}$ |
| $I_o$ | Peak Intensity | $n_o c \dfrac{|A_o|^2}{2\pi}$ |
| $L_{mp}$ | Multi-Photon Absorption | $\dfrac{1}{\beta^m I_o^{(m-1)}}$ |
| $\rho_o$ | Normalization constant. Representing the peak plasma absorption rate in normalized units. | $\dfrac{\beta^m I_o^m \tau_p}{m\hbar\omega}$ |
| $\rho$ | Reduced plasma density | $\dfrac{\rho_e}{\rho_o}$ <br> $\rho_e$ is the electron density |
| $L_{pl}$ | Multi-Photon Absorption Length[7] | $\dfrac{2}{\rho_o \sigma \omega \tau_c}$ |
| $\alpha$ | Avalanche Ionization Coefficient | $\dfrac{\sigma I_o \tau_p}{n_o^2 E_g}$ |

Appendix A Table 2: A list of constant symbols of Eq. (1). The defining equations of each constant is listed.

## Appendix B: Initial Sampling Conditions and Adaptive Step Size Algorithm

This subsection will now proceed in deriving initial sampling requirements and show the adaptive step-size algorithm used in the simulation. Firstly, the derivation of the initial sampling intervals will be shown.

The above GNLSE is only valid for a reduced angular frequency range (inverse angular variable of $\tau$) from $-0.5\omega_o\tau_p < w < 0.5\omega_o\tau_p$ due to the slow-varying approximation. This means, a bandlimited approach can be assumed and thus, the Nyquist criterion can be applied to calculate the spacing in $\tau$ needed. For the FFT algorithm used, sampling is at the Nyquist Criterion.

Accordingly, $d\tau$ the spacing in $\tau$ is:

$$d\tau \leq \frac{2\pi}{2(0.5\omega_o\tau_p)} = \frac{2\pi}{\omega_o\tau_p} \tag{15}$$

Where, $\omega_o$ corresponds to the central frequency of the initial input pulse. The range of $\tau$ is set to a desired maximal range ($\Delta\tau$-user specified) and then $dw$, in angular radian units is calculated as:

---

[7] This term was modified from the original term. Due to a typographic error, the formula was originally incorrect (units did not match).

$$dw \leq \frac{2\pi}{2(0.5\Delta\tau)} = \frac{2\pi}{\Delta\tau} \quad (16)$$

The MATLAB FFT algorithm employed in this simulation defines window sizes at the equality condition for the above two equations. The FFT algorithm is periodic in nature and relies on a Fourier series representation of the function in the window of a domain. The arrays are designed such that the window size is an integer multiple of the spacing and that there is an even amount of array elements. The even condition insures that the matrix swapping needed in the MATLAB FFT algorithm does not introduce element swapping error. The positive endpoints value of the windows is reduced in magnitude by one step size in relation to the negative endpoints magnitude due to these constraints of the window, which also satisfies the periodic nature of the FFT. As well from the integer multiple condition, the zero frequency is always sampled. If these conditions are satisfied in one domain, they are automatically satisfied in the inverse (frequency) domain. These conditions assure that the frequency domain value (and vice-versa, this also applies for the time domain when taking the inverse FFT) corresponding to the frequency array element number is unambiguous. For example, the frequency domain value is the value obtained from incrementing with the spacing in Eq.(16) from the minimal frequency-$0.5\omega_o\tau_p$.

The amount of data points for both the frequency range and the time range are the same:

$$Time\ array\ size = \frac{\Delta\tau\omega_o\tau_p}{2\pi} \quad (17)$$

For the Spatial resolution there are no imposed upper bounds in momentum. The upper bound in momenta is introduced from the maximum frequency bound and using the paraxial approximation. The Nyquist criterion for the normalized angular $k_\chi, k_\psi$ momenta, reads as:

$$|k_\chi, k_\psi| \leq \frac{nS_p(1.5\omega_o)}{c\tau_p} \quad (18)$$

$n$ is the refractive index value at $1.5\omega_o$. Due to the intensity dependent nature of the index of refraction the maximal momenta can be higher than this estimation.
As in the case for the temporal range, the spatial range is a free user parameter in the model. The normalized spatial step size and the reduced angular momenta step sizes can be defined in the same way as derived above for the normalized time-reduced angular frequency case.

$$d\chi, \psi \leq \frac{1}{2\left(\frac{1.5\omega_o nS_p}{c\tau_p}\right)} = \frac{2\pi c\tau_p}{3(n\omega_o S_p)} \quad (19)$$

The range of $\chi, \psi$ is set to a desired domain length and then $dk_{\chi,\psi}$ are calculated as:

$$dk_{\chi,\psi} \leq \frac{2\pi}{2(0.5\Delta\chi,\psi)} = \frac{2\pi}{\Delta\chi,\psi} \tag{20}$$

As with the time and frequency array size, the space and momentum's two-dimensional array size are the same:

$$Space\ array\ size = \left(3\Delta\chi,\psi \frac{n\omega_o S_p}{2\pi c\tau_p}\right)^2 \tag{21}$$

At the start, the spatial and temporal grid sizes should be chosen such that the input signal decays to zero before the edges of the window. However, this is not a strict requirement given the use of the adaptive step-size algorithm described below. The adaptive step-size algorithm can be used to account for the expanding domain windows and the changes in the required sampling increments and thus avoid aliasing errors.

### B.1 Numerical Example from Simulation

The sampling ranges and step-size for all coordinates for the simulation at numerical convergence for the example YAG system is given in Table 4. The self-steepening term requires a large number of sampling points. Consequently, convergence is obtained with time step sizes representing a bandwidth that would extend into unphysical negative real frequencies. Thus, zero padding was placed at the (negative) reduced frequency that would correspond to the zero frequency and after. This provides the small time steps required while not violating the underlying physics. On the positive reduced frequency side, zero padding was placed after the reduced frequency that corresponds to the wavelength of 1700nm as this is where the region of validity of the simulation ends (due to the zero dispersion point of YAG). This interpolation is necessary because the self-steepening compression and expansion effect on the temporal envelope function depends on the difference of two neighbouring intensity points. However, the amplitude reorganization needed to ensure that the overall self-steepening term is unitary relies on the derivative of the intensity. Therefore, for the derivative term to "balance out" the expansion and compression effect of the intensity difference of the neighbouring points, the time separation of these points must be on a timescale where the derivative becomes relevant and representative. Meaning at a timescale where the intensity difference converges with the derivative multiplied into the time interval. Finally, this translates to the two neighbouring intensity points being a differential distance away from each other in the time coordinate. Which numerically translates to the sampling interval being much more constrained for the self-steepening to remain unitary. *To summarize,* the self-steepening effect has additional constraints on the sampling interval for it to be energy conservative and to remain physically valid. This would translate into a maximal lower frequency that would be negative. The large padding ensures that aliasing wrap-around effects are not present.

| Simulation Parameter | Value |
|---|---|
| Normalized Time-step value | 0.0251 |
| Lower normalized time window value | -5.0187 (corresponds to 25.595 fs)[8] |
| Upper normalized time window value | 4.9936 (corresponds to -25.467 fs) |
| Lower reduced frequency | -125.190 |
| Upper reduced frequency | 124.57 |
| Minimum *simulated* reduced frequency | -30.6730 |
| Maximum *simulated* reduced frequency | 31.2990 |
| Reduced frequency step | 0.620 |
| Normalized transverse spatial step | 0.0258 |
| Upper normalized transverse space value | 4.9788 (corresponds to -74.682 µm) |
| Lower normalized transverse space value | -5.0046 (corresponds to -75.068 µm) |
| Lower reduced momentum | -121.7826 |
| Upper reduced momentum | 121.1548 |
| Reduced momentum step | 0.6277 |
| Normalized propagation step size | 0.0082 |
| Maximum normalized propagation value (corresponding to 2 mm) | 4.9268 |
| Space and momentum grid | 2D rectangular grid, 2D FFT Algorithm used |
| Time and frequency grid | 1D array, 1D FFT algorithm used |
| Time and frequency array size | 400 |
| Space and momentum array size | $388^2 = 150{,}544$ |
| Total Array Elements in a propagation step | 60,217,600 |

Table 4: Numerically convergent parameters for the example system shown in the results section. The propagation step size converged to a value corresponding to the central wavelength which indicates that the paraxial approximation is justified.

The above simulation was performed on a workstation with 20 cores, requiring 16 gb of RAM. The computation time per propagation step without saving data is 45 seconds yielding a total time of 7.5 hours for 600 propagation slices.

**B.2 Sampling Criteria and the Adaptive Sampling Step-Size Algorithm**

In this section grid size considerations will be considered to reduce global step-size error. Provided the Nyquist criterion for the sampling intervals is satisfied for the original input pulse, the error originates from: Under-sampling the instantaneous phase variation contributions from the exponential operators, the exponential error due to the real exponential terms in the operator, commutation error between operators, and error due to the mean-value approximation used. The appropriate Nyquist criterion when applied to the phase terms is sufficient to subdue the phase error making this method, through its spectral nature extremely precise. The real exponential error can be reduced in a similar way: By considering characteristic lengths of these exponential decaying terms. This will be rigorously derived below. The commutation error is dependent on the ordering of how the operators are applied (the symmetrisation). This error is reduced by numerically experimenting with the ordering of the three operators.

In general, acquiring a proper upper bound calculation for the longitudinal step size is rather difficult: Unless, the mean field "slow-varying" approximation can quantitatively be defined. This would involve a numerical recursion scheme. Physical properties of the system being studied can help [252525]. For example, [2626] derives longitudinal step size conditions based on commutation relations and uncertainty relations between operators. For the mean field error, simple convergence by varying the propagation coordinate ensured reduction of this error.

---

[8]The normalized time is in fact the reversed normalized time. The time window is reversed for the MATLAB simulation due to the engineering definition used for the Fourier transform in MATLAB. The equation was rewritten for a time reversed normalized time coordinate. While, the frequencies could have been reversed instead, this would be a less elegant approach as it would translate to more overall computational operations in the overall simulation program.

There are two main topics to consider when defining the step size for the domains of $u$:

1) The step size should be appropriate such that the exponent terms do not vary faster than the Nyquist criterion defined for the system (otherwise there could be under-sampling errors that iteratively grow) producing aliasing effects and low sampling resolution effects.

2) Under most cases a good first estimate of propagation step size corresponds to the inverse of the highest ratio of coefficients in Eq. (1) (i.e., $\max\left[\frac{L_{df}}{L_{nl}}, \frac{L_{df}}{L_{mp}}, etc\right]$).

At the start of the simulation sampling is at or below the Nyquist criterion for the input pulse. The propagation step size is calculated from point 2). If, however, the step sizes need to be varied, the simulation parameters are updated accordingly. The variation algorithm will be rigorously described in the subsections below starting with the Nyquist sampling conditions of the FFT algorithm employed.

**B.3 The Adaptive Step Size Algorithm**

While the frequency range is limited by the slow varying approximation, extending the frequency ranges can still yield insight to the system response over the increased bandwidth. The results section demonstrates that the simulation can still fit experimental results over frequency ranges that violate the slow-varying approximation. Window sizes need to be updated due to the various non-linear and linear effects. For example, the temporal and spatial window size should be increased, due to the GVD walk off the optical pulses and diffractive expansion in the transverse spatial coordinates. The following method offers a rigorous adaptive algorithm to adjust the original Nyquist Criteria. This adaptive split-step method has not been completely implemented in the simulation due to the computation resources needed and because it is not needed for the simulation as applied to the example system in this paper. However, physical insights can be derived from the analysis that goes into the adaptive step size algorithm, especially when dealing with the $\hat{C}$ operator below. The potential for its application and the physical insights it yields justifies presenting it in this paper. The rigorous numerical implementation of this algorithm will be presented in a follow up publication.

**B.3.1 Phase Contributions: Part of Algorithm that Evaluates the Effects of the First Derivative**

Not only is it important to calculate the original step-size from the Nyquist conditions of the system but also to factor the additional instantaneous phases from the operators. In the respective domains, the operators add instantaneous phase (i.e., frequencies) that translates to higher maximal values in the inverse domain. For example, considering the self-phase modulation term in time translates to a broadening of the frequency domain. To evaluate the new domain boundaries a checking algorithm is employed in this computational method. The exponential operators can have both imaginary and real arguments in respective domains. The derivative of the imaginary arguments w.r.t to the domain being considered at values in the domain yields the additional instantaneous phase contribution at that domain value. For real arguments, the negative terms and positive terms are considered differently. For the negative terms, it is assumed that the derivative w.r.t to the domain being considered at a domain value yields the characteristic length of the decaying exponent at a domain value. From this, new sampling conditions are obtained at every propagation slice and verified with the original.

In Appendix B.3.1 and B.3.2, only the imaginary argument terms of the exponential $\hat{A}$, $\hat{B}$ and $\hat{C}$ operators are considered. All derivations in these subsections are over the imaginary argument terms of these exponential operators. First, when considering the phase properties of $u$ in a specific domain, operators who are applied in that domain are grouped together, irrespective of other domains they are applied in. This is how the algorithm starts. This is because, as will be seen below, the operators phase function in a specific domain gives easy to access information on its impact on the phase of $u$ in that domain. The algorithm calculates the maximal phase variation imposed on $u$ from the operator phase functions at the end of the propagation slice (here labelled as slice $k$).

The inverse domain to the domain being considered is labelled as the 'frequency' domain (even if physically this is not the case: I.e., the time or space domain). The instantaneous phase function, $\frac{\partial \emptyset_k}{\partial x}$ ($x$ representing a domain variable) at a domain value, can be viewed as the 'central frequency', of the 'frequency' representation of a portion of the amplitude function of $u$ around that domain value. If the Nyquist sampling condition is violated, aliasing to lower 'frequencies' will occur. When this violating 'frequency' occurs, the representation of the corresponding portion of the amplitude function in the 'frequency' domain will wrap around to the other end of the inverse domain window because its 'central frequency' will be placed there. To prevent the aliasing effect, sampling step-sizes are adjusted using the Nyquist criterion for the domain obtained from the maximal instantaneous phase. The domain step-size must satisfy:

$$dx_k \leq \frac{2\pi}{2\boldsymbol{max}\left[\boldsymbol{abs}\left[\frac{\partial \varphi_k}{\partial x}\right]\right]} \tag{22}$$

Max is calculated over the full set of values ($K_\chi, K_\psi, w$). If the original sampling interval increment, $dx_0$, in the respective domain is smaller or equal to this updated sampling interval increment than there is no need to adjust the sampling interval in the domain being considered. The window size of the inverse domain is:

$$\Delta x^{-1}{}_k = 2\frac{2\pi}{2dx_k} \geq \boldsymbol{max}\left[\boldsymbol{abs}\left[\frac{\partial \varphi_k}{\partial x}\right]\right] \tag{23}$$

$x^{-1}$ indicates an inverse domain (i.e, the temporal domain for frequency and spatial for momentum).

To guarantee that the inverse domain length is adequate, $dx_k$ is always taken to be smaller than the equality condition in Eq. (22). As well, a value is chosen such that the even array and integer multiple conditions of the domain window to the step size (described in subsection B.2) is maintained.

To simplify the analysis, the derivation of the algorithm first starts with the w domain and it is assumed that only the exponential $\hat{A}$ operator acts in this domain ($\hat{C}$ is omitted for the time being). Also, since $\hat{A}$ acts over the ($K_\chi, K_\psi, w$) dimensions, these dimensions share a common operator phase function through $\hat{A}$ and must be considered together. The total phase derivative accumulated at the end of a slice $k$ just considering the $\hat{A}$ operator is given as:

$$\frac{\partial \varphi_k}{\partial x} = \frac{\partial \varphi_0}{\partial x} + \sum_{n=1}^{n=k} \frac{\partial \varphi_A}{\partial x} \qquad (24)$$

$x$ is a place holder for the domain being considered in the three dimensional set $\varphi_A$ stands for the imaginary arguments of the exponential operator $\hat{A}$ (the step size increment in the propagation coordinate is included as a coefficient within the imaginary argument), $\varphi_0$ is the original phase of the input pulse. There are no other operators that are applied in the $K_\chi, K_\psi$ domains. This means that there are no other operators added in the above summation when $x = K_\chi, K_\psi$.

From Eq. (24), it can be seen that the window size in the inverse domain (barring contributions from the exponential $\hat{C}$ operator explained below) can be calculated before the start of the simulation since the exponential $\hat{A}$ operator only relies on the frequency, momentum range and propagation step size and does not rely on the $u$ inputted into the iteration. As an illustrative example, closed form expressions will be derived for the spatial and temporal window size that accounts for dispersive and diffractive effects from the exponential $\hat{A}$ operator omitting the first term in Eq. (24) (thus, in the case of a transform limited input pulse). This is a lower bound calculation since it does not factor in the $\hat{B}$ and $\hat{C}$ operator effects. Since, the $\hat{B}$ operator is not applied in the domains of Eq. (24), its effects are not readily calculable and lies out of the scope of the algorithm (except for the discussion in section B.3.2). The phase function derivative of the exponential $\hat{A}$ operator is the same as the derivative of the $\hat{A}$ operator (omitting the imaginary constant coefficient) since this unitary operator has no real exponential arguments. The derivative of its phase function is given as:

$$\frac{\partial \varphi_A(k_\chi, k_\psi, w)}{\partial w} = -\frac{1}{4\omega_o \tau_p}\left(1 + \frac{1}{\omega_o \tau_p} w\right)^{-2}(k_\chi^2 + k_\psi^2) + 2\frac{L_{df}}{L_{ds}} w \qquad (25)$$

$$\frac{\partial \varphi_A(k_\chi, k_\psi, w)}{\partial K_{\chi,\psi}} = -\frac{1}{4}\left(1 + \frac{1}{\omega_o \tau_p} w\right)^{-1}(2K_{\chi,\psi}) \qquad (26)$$

Since Eqs. (25), (26) do not depend on the propagation coordinate, the total instantaneous phase w.r.t $(K_\chi, K_\psi, w)$ at a slice $k$ for a transform limited input can be approximated using Eq. (24) as:

$$\frac{\partial \varphi_k}{\partial x} = \sum_{n=1}^{n=k} \frac{\partial \varphi_A}{\partial x} = \frac{\partial \varphi_A}{\partial x} \Delta\varsigma$$

Where, $\Delta\varsigma$ is the length of the crystal in the reduced propagation coordinate. Using the equality condition for Eq. (23) the following can be obtained for the end slice of the crystal:

$$\Delta\tau = 2\mathbf{max}\left[\mathbf{abs}\left[\frac{1}{4\omega_o\tau_p}\left(1+\frac{1}{\omega_o\tau_p}w\right)^{-2}(k_\chi{}^2+k_\psi{}^2)+2\frac{L_{df}}{L_{ds}}w\right]\right]\Delta\varsigma \qquad (27)$$

$$\Delta\chi,\psi = 2\mathbf{max}\left[\mathbf{abs}\left[\frac{1}{2}\left(1+\frac{1}{\omega_o\tau_p}w\right)^{-1}(K_{\chi,\psi})\right]\right]\Delta\varsigma \qquad (28)$$

Eq. (27) and (28) give an estimate lower bound window size estimate in the temporal and spatial domains to account for dispersion and diffraction.

From the operator domains of application, the domains are grouped together as $(K_\chi, K_\psi, w)$, $(\chi, \psi, \tau)$, $(\chi, \psi, \tau, w')$. When considering the domain subsets, the $\hat{C}$ operator introduces an additional order of complexity due to its 4 domain nature. The exponential $\hat{C}$ operator imposes phase on the frequency representation and on the $(\chi, \psi, \tau)$ as will be seen. Firstly, its contribution of instantaneous phase in the frequency domain will be derived. Eq. (24) is updated as follows for the frequency derivative:

$$\frac{\partial\varphi_k}{\partial w} = \frac{\partial\varphi_0}{\partial w} + \sum_{n=1}^{n=k}\left(\frac{\partial\varphi_A}{\partial w} + f_n(\tau,\chi,\psi)d\varsigma\right) \qquad (29)$$

Where, is the coefficient of $(-iw')$ in the $\hat{C}$ operator exponential argument. $d\varsigma$ is the propagation step and is explicitly shown for the $\hat{C}$ operator term. This accounts for the maximal time delay generated by the group velocity dependence to the intensity varying refractive index which is described in the self-steepening effect and the exponential $\hat{C}$ operator. Thus, Eq. (29) calculates the temporal window from both the linear dispersion and from the group velocity effects of the self-steepening term. Since, $\hat{C}$ does not act in the momentum domains, this additional term is not considered (i.e., the $\hat{C}$ exponential operator phase function derivative is zero w.r.t to $k_{\chi,\psi}$).

The rest of the effects of the exponential $\hat{C}$ operator acts on $(\chi, \psi, \tau)$ and its instantaneous phase contribution will now be considered in that domain set. The $\hat{C}$ operator is grouped with the $\hat{B}$ operator. The phase derivative is given as (from section 4):

$$\frac{\partial\varphi_C}{\partial\tau} = \frac{\partial\varphi_0(\chi,\psi,G)}{\partial G}\left[1 - \frac{\partial f(\tau\chi,\psi)}{\partial\tau}d\varsigma\right] \qquad (30)$$

For the χ, ψ domains the phase derivative is:

$$\frac{\partial\varphi_C}{\partial\chi,\psi} = \left[\frac{\partial\varphi_0(\chi,\psi,G)}{\partial\chi,\psi}\right]_G + \left[\frac{\partial\varphi_0(\chi,\psi,G)}{\partial G}\frac{\partial f(\tau\chi,\psi)}{\partial\chi,\psi}d\varsigma\right] \qquad (31)$$

Numerically, the above equations are not used. Instead the phase slices in time are reorganized by the offset term $f(\tau, \chi, \psi)d\varsigma$ shown in section 4 and then the temporal and spatial numerical derivative is calculated.

For the sake of simplicity, the order of how the exponential $\hat{B}$ and $\hat{C}$ operators are applied is omitted from the analysis and $u$ inputted into the slice is used for the exponential $\hat{C}$ operator calculation. The instantaneous phase in $(\chi, \psi, \tau)$ can be estimated as:

$$\frac{\partial \varphi_k}{\partial x} = \frac{\partial \varphi_0}{\partial x} + \sum_{n=1}^{n=k}\left(\frac{\partial \varphi_B}{\partial x} + \frac{\partial \varphi_C}{\partial x}\right) \tag{32}$$

$\frac{\partial \varphi_B}{\partial x}$ is the derivative of the exponential $\hat{B}$ operator imaginary argument (the propagation step size for the exponential $\hat{B}$ and $\hat{C}$ operator is factored into the argument and not explicitly shown here). In this case, the exponential $\hat{B}$ operator term in Eq. (32) can be evaluated by just considering the exponential arguments of the operator (updated with $u$ outputted from the previous slice). However, the exponential $\hat{C}$ operator term cannot be evaluated in such a manner. The original phase distribution must be reordered going into each propagation slice due to the reordering effect of the exponential $\hat{C}$ operator. From this algorithm one can deduce the new bandwidth of the momentum domains as well as the frequency domain ($w$). For the sampling condition, Eq. (22) applies.

**B.3.2 Additional Considerations: The Second Derivative of the Phase**

Due to the frequency generation from the intensity variant refractive index, not only does the bandwidth in the frequency domains need to be extended, as shown above, but the new resolution in the angular frequency domains change (momentum, time-frequency). The angular frequency step-size needs to be updated. As already stated, the intensity variant phase derivative obtains the instantaneous frequency. As used above, the maximum first derivative obtains the maximum frequency. Therefore, the minimum of the second derivative w.r.t the inverse domain of question, will obtain the minimal change in frequency. This will directly determine the new sampling step size in the inverse frequency domains: The step-size must be at this minimal frequency change or below. The incoming phase of $u$ need not be considered here, as the minimum difference between frequencies is all that is important. However, additional constraints must be considered (Figure 8).

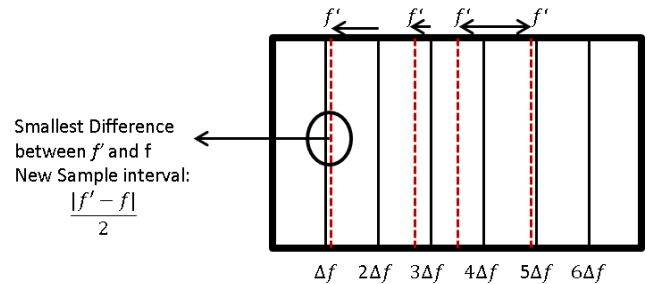

Figure 8: Not only does the minimal frequency variation must be considered, but the absolute difference between the new frequencies and the incoming frequency grid as depicted in this diagram. This consideration yields Eq.(33). The factor of half is maintained to insure adequate sampling.

With these additional constraints, for the $\hat{B}$ and $\hat{C}$ exponential operators, in the $(\chi, \psi, \tau)$ domains the grid spacing in the inverse domains must satisfy:

$$df_{new} \leq 0.5\mathbf{min}\left[\mathbf{min}\left[\mathbf{rem}\left(\frac{\frac{\partial^2 \varphi_B}{\partial x^2}dx}{df}\right), df - \mathbf{rem}\left(\frac{\frac{\partial^2 \varphi_B}{\partial x^2}dx}{df}\right)\right], \mathbf{min}\left[\mathbf{rem}\left(\frac{\frac{\partial^2 \varphi_B}{\partial x^2}dx}{df}\right), df - \mathbf{rem}\left(\frac{\frac{\partial^2 \varphi_B}{\partial x^2}dx}{df}\right)\right]\right] \quad (33)$$

In Eq.(33), the propagation step size is factored into the derivative terms. dx is the domain step-size being considered. $df$ is the original frequency step (inverse domain of *x*). Rem is the remainder function evaluated at all coordinate value, it is a 3-D array. The inner mins are also 3-D arrays. The outer min represents the operation of evaluating the minimum of the two global mins for the two inner mins. It is evaluated over two scalars. This is always computed below the equality condition.
The last consideration is that due to the Fourier series nature of the FFT, the minimal spacing calculated in Eq.(33) must be a divisor of the maximal frequency. A spacing is chosen that is less than the upper bound in Eq.(33) but satisfies the integer multiple criterion and the even array criterion. Going back to the beginning of Appendix B.3.1, in discussion of the exponential $\hat{A}$ operator and $\hat{C}$ operator for the $(K_\chi, K_\psi, w)$ domains, the second derivative there to is evaluated, exactly in the same manner as the above analysis. Physically, time and space shifts from the linear GVD and spatial diffraction can shift in intervals smaller than the grid spacing causing errors in the inverse space.

### B.3.3 Putting it all together

The minimum step-sizes calculated from the above first derivative and second derivative contributions are then used for all respective domains. Consequently, the second derivative step-size of the exponential $\hat{B}$ and $\hat{C}$ operator is compared with the first derivative step-size of the exponential $\hat{A}$ and $\hat{C}$ operator and the minimal value is used. The exponential $\hat{A}$ and $\hat{C}$ operators second derivative step-size is compared to the first derivative step-size of the exponential $\hat{B}$ and $\hat{C}$ operator and the minimal value is used.
Once the optimal window sizes are found, $u$ is interpolated with these new window sizes. This means that once the lower bound window is found, a new window size is chosen in accordance to what was already described. The $u$ array is padded by zeroes equally on both the negative and positive domain side to achieve the desired window size. However, after this procedure, the $\hat{B}$ and $\hat{C}$ operators change. For that reason, the procedure to calculate window sizes must be carried out again. This is done until convergence to a set value for each window is found. Once $u$ is accordingly padded, $u$ is recalculated at the last propagation slice (now beginning of slice *k*) and the procedure is carried on iteratively with propagation slice.

### B.3.4 Effects of the Negative Exponential Arguments in the Algorithm

For negative real exponential arguments, the decaying exponential argument should not go to $e^{-1}$ within the step-size. To insure this, the sampling interval should be half the value of the exponential decay length. This is given as the maximum coefficient value of the real exponential arguments of operators for the domain that is being considered. In terms of the inverse domains, a real exponential argument occurs only for the w domain due to the exponential $\hat{C}$ operator. Thus, at slice *k* the characteristic decay length associated with this argument is:

$$L_{dw}^{-1} = [f(\tau, \chi, \psi)]\Delta\varsigma \qquad (34)$$

Therefore, the new sampling condition is calculated as:

$$dw_k = \frac{2\pi}{2\mathbf{max}[L_{dw}^{-1}]} \qquad (35)$$

For the w, the original sampling step-size should be smaller or equal to the minimum between the imaginary and real calculated values:

$$dw_k = \mathbf{min}[dw_{k,i}, dw_{k,R}] \qquad (36)$$

For the exponential $\hat{B}$ operator the same analysis can be carried for the $(\chi, \psi, \tau)$. It is difficult to assess the effects of the real exponential $\hat{C}$ operator arguments in $(\chi, \psi, \tau)$: While the coefficient function of this argument is over $(\chi, \psi, \tau)$, it effectively reshapes the *amplitude* in the $w$ domain which is not as simple as introducing phase delays in the $w$ domain (as in the imaginary argument case).
To further the discussion, the step-size in the propagation coordinate introduces a linear scaling for all phase and decaying exponential effects described above. Thus, by reducing this step-size the onus of tuning other domain step-sizes is reduced.

## Appendix C: Relevant Omissions, singularities and extensions to the GNLSE

This appendix will cover certain omissions and errors in Eq. (1) as well as extensions to the equation. Eq. (1) is linearized in such a manner that higher order terms in the Taylor expansion about a given coordinate point $(\chi, \psi, \tau, \varsigma)$ are omitted. Therefore, the equation can be written in terms of the constants stated in Appendix A. Also, it is assumed that the original pump peak contributes the most to the generated white light and thus, the constants are only calculated for the original pump signal. The slow-varying approximation of the envelope imposes the condition that the bandwidth is equal to the central frequency of the pump. In fact, mathematically, the equation can be extended past this bandlimited requirement. This was done to simulate the experimental results. Mathematically, the only constraint is that the frequency bandwidth cannot violate the series convergence to the functional form of $\hat{A}$ in its inverse space. Otherwise, the full divergent series must be considered, yielding un-physical solutions. The series will be divergent if:

$$\mathbf{abs}\left[\frac{1}{\omega_o \tau_p} w\right] \geq 1$$

The series obtained from the binomial expansion of the inverse coefficient w.r.t to the time derivative is represented in the frequency domain and then a closed form expression obtained from its convergence is obtained to represent the $\hat{A}$ operator in the frequency domain. However, the series is divergent at the positive frequency limit and the closed form expression then cannot be obtained. Therefore this limit is necessary[9]. The negative limit is obvious: both the closed form expression and thus the series are divergent due to the expression going to zero in the denominator. The frequency representation of $u$ has to be bandlimited, i.e., go to zero at these endpoints, to ensure that when the series becomes invalid, the overall product between the series and $u$ is zero. Due to the above limit, the non-zero padded simulated bandwidth can at-most be extended to :

$$-\omega_o \tau_p < w < \omega_o \tau_p$$

---

[9] However, it can be seen from how Eq. 1 is derived that the positive limit can be omitted. It can be shown that an alternate derivation [11] of Eq. 1 is solely carried out in the frequency domain and derives the closed form

If the original slow-varying approximation used in the NLSE is relaxed.

Other errors are more subtle. For example, as more frequencies are generated, the characteristic lengths of the terms of Eq. (1) change and do not necessarily equal the original. This is because they are functions of the all functions of the instantaneous frequency $\omega(\chi, \psi, \tau)$ and indirectly functions of $\chi, \psi, \tau$.

For example, the $L_{mp}$ length changes since the order of multi-photon ionization and the coefficient of multi-photon ionization ($\beta^m$) changes with frequencies. At higher frequencies away from the central frequency the equation underestimates the absorption to the plasma. However, it is assumed that the intensity of these components, especially near the end of the spectral bandwidth are much lower than the pump and the absorption is thus, in this regard, negligible compared to the pump.

To minimize such errors in $L_{mp}$, an expression using the instantaneous frequency can be derived. However, to be accurate, Eq. (1) would have to be re-derived without the slow-varying approximation. $L_{mp}$ can be calculated using $u$ coming from the end of the previous slice, and redefined as:

$$L_{mp}(\chi, \psi, \tau) = \frac{1}{\beta(\omega(\chi, \psi, \tau))^m I_o^{m-1}}$$

$$\omega(\chi, \psi, \tau) = \mathbf{abs}\left[\omega_O + \frac{\partial \emptyset(\chi, \psi, \tau)}{\partial \tau}\right]$$

$\emptyset$ is the phase of $u$, $I_o$ is the original peak intensity of the input pump pulse. $\omega(\chi, \psi, \tau)$ is the instantaneous frequency of the chirped pulse, $m$ is the order corresponding to $\omega(\chi, \psi, \tau)$ ($m$ is as well a function of $\chi, \psi, \tau$). Here it is considered that only the instantaneous phase of the pulse would register as a frequency change for the $\beta^m$ coefficient. The above derived approximation assumes that the overall $L_{mp}$ is a slowly varying function such that its $\tau$ derivative is negligible. Therefore, the derivative of $L_{mp}$ can be omitted when setting up the operators. If this assumption is violated the $\widehat{B}$ operator would have to be changed to account for the additional derivative term; this also holds true for all other characteristic lengths as well.

Continuing with the assumption that the characteristic lengths vary slowly across the time and space coordinates, $L_{ds}$ could be updated in a similar manner as $L_{mp}$ whereby its values are taken at the instantaneous frequency. $L_{df}$ would be taken at the instantaneous momentum corresponding to the instantaneous frequency for the time step and the assumption is used that the spot size used in the calculation is the same as the original pump signal. $L_{nl}$ can be updated in a similar manner.

---

expression without use of a series convergence in the frequency domain. Then the $\widehat{A}$ operator is only divergent when the denominator goes to zero, so when the negative limit occurs.

Finally, the equation is a scalar equation valid for isotropic material where the white light is totally depolarized or linearly polarized. Under the circumstance that it is depolarized, one can assume that the individual polarization states see the same average $u$ and thus, this equation models the propagation of this average $u$. The simulation can easily be extended to consider polarization. There will be two scalar coupled equations. The method is applied to each of those equations.